\documentclass[aps,twocolumn,superscriptaddress,nobibnotes]{revtex4-2}

\usepackage{mhchem}

\usepackage{graphicx}
\usepackage{caption}

\usepackage{siunitx}
\DeclareSIUnit{\angstrom}{\textup{\AA}}
\usepackage{hyperref}
\hypersetup{colorlinks=true, citecolor=blue, urlcolor=blue, linkcolor=blue}

\begin{document}
\title{Multiple origins of extra electron diffractions in fcc metals}

\author{Flynn Walsh}
\thanks{equal contribution first author}
\affiliation{Materials Sciences Division, Lawrence Berkeley National Laboratory, Berkeley, CA, USA}
\author{Mingwei Zhang}
\thanks{equal contribution first author; corresponding author}
\email{mwwzhang@ucdavis.edu}
\affiliation{Materials Sciences Division, Lawrence Berkeley National Laboratory, Berkeley, CA, USA}
\affiliation{National Center for Electron Microscopy, Lawrence Berkeley National Laboratory, Berkeley, CA, USA}
\affiliation{Department of Materials Science \& Engineering, University of California, Berkeley, CA, USA}
\affiliation{Department of Materials Science \& Engineering, University of California, Davis, CA, USA}
\author{Robert O. Ritchie}
\affiliation{Materials Sciences Division, Lawrence Berkeley National Laboratory, Berkeley, CA, USA}
\affiliation{Department of Materials Science \& Engineering, University of California, Berkeley, CA, USA}
\author{Mark Asta}
\affiliation{Materials Sciences Division, Lawrence Berkeley National Laboratory, Berkeley, CA, USA}
\affiliation{Department of Materials Science \& Engineering, University of California, Berkeley, CA, USA}
\author{Andrew M. Minor}
\thanks{corresponding author}
\email{aminor@berkeley.edu}
\affiliation{Materials Sciences Division, Lawrence Berkeley National Laboratory, Berkeley, CA, USA}
\affiliation{National Center for Electron Microscopy, Lawrence Berkeley National Laboratory, Berkeley, CA, USA}
\affiliation{Department of Materials Science \& Engineering, University of California, Berkeley, CA, USA}

\begin{abstract}

Diffuse intensities in the electron diffraction patterns of concentrated face-centered cubic solid solutions have been widely attributed to chemical short-range order, although this connection has been recently questioned.
This article explores the many non-ordering origins of commonly reported features using a combination of experimental electron microscopy and multislice diffraction simulations, which suggest that diffuse intensities largely represent thermal and static displacement scattering. A number of observations may reflect additional contributions from planar defects, surface terminations incommensurate with bulk periodicity, or weaker dynamical effects.

\end{abstract}

\maketitle

\section{Introduction}

High-entropy alloys were originally conceived as crystalline solid solutions with effectively ideal configurational entropies, i.e., complete compositional disorder. Over the past decade, however, it has become increasingly apparent that alloys of multiple principal elements can contain chemical short-range order (SRO) among neighboring atoms \cite{walsh23}.
While there exists little agreement on the nature, extent, or relevance of the ordering realized in most systems, the \textup{\AA}-scale chemical structure of concentrated alloys remains intensively investigated as a tantalizing scientific question with potential implications for the unprecedented damage tolerance of materials such as CrCoNi \cite{liu22,yu22,kireeva23}.

In the latter half of the last century, SRO in binary alloys was commonly characterized using the diffuse scattering of monochromatic x-rays or neutrons, which can semi-quantitatively characterize mean-field chemical environments \cite{ice99}, although the interpretation of measurements becomes rapidly more challenging with increasing compositional complexity \cite{schonfeld19}.
Moreover, these methods require specialized synchrotron experiments and the preparation of single-crystalline samples.
The majority of contemporary characterization efforts have thus preferred to analyze diffraction patterns (DPs) obtained from transmission electron microscopy (TEM), which can spatially resolve local features in two dimensions in a manner that may be complemented by techniques such as strain or composition mapping.

Most of this work has examined face-centered cubic (fcc) alloys containing several 3\textit{d} transition metals as principal elements, which are the subject of this article.
While structural specifics may vary among systems, the reduction of unfavorable V-V \cite{kostiuchenko19,chen21} and Cr-Cr \cite{schonfeld19,ghosh22} nearest neighbor pairs is generally expected, as may largely be explained by electrostatic interactions arising from interatomic charge transfer \cite{papez23}.
The kinetics of this type of ordering are not well established, but recent results suggest that SRO can at least initially form quite rapidly \cite{du22,teramoto23}, such that a significant degree may be effectively ubiquitous.

\begin{figure*}
\includegraphics[width=\textwidth]{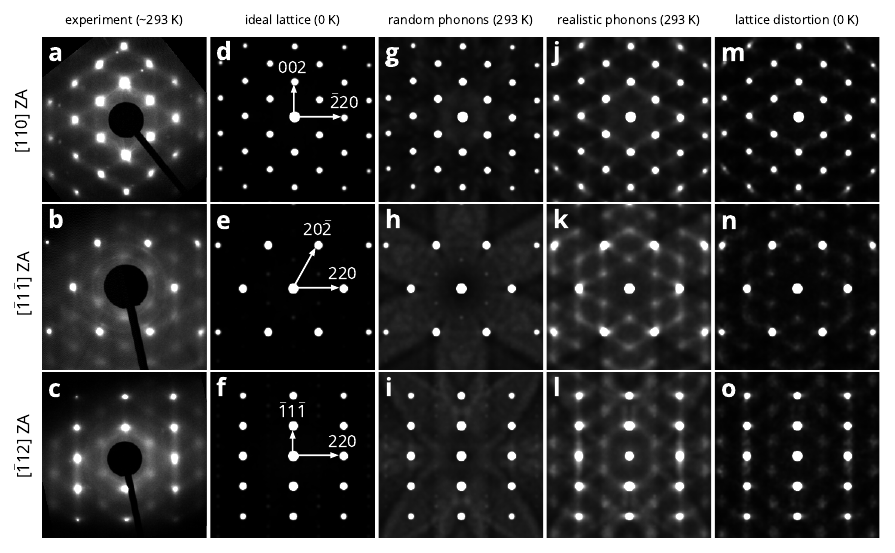}
\caption{\label{fig:crconi}
    \textbf{Extra diffractions from thermal and static displacement scattering in CrCoNi.} Selected-area electron DPs of CrCoNi containing (a) streaking in the [110] ZA, (b) diffuse $\frac{1}{3}\{422\}$ intensities the $[\bar{1}1\bar{1}]$ ZA, and (c) diffuse $\frac{1}{2}\{311\}$ intensities in the $[\bar{1}12]$ ZA. Coordinates are given in (d-f), which provide equivalent DPs calculated for ideal CrCoNi lattices. (g-i) demonstrate the diffuse scattering expected from random thermal displacements, while (j-l) show calculations for more realistic molecular dynamics trajectories, which largely reproduce experiment in the $[\bar{1}1\bar{1}]$ and $[\bar{1}12]$ ZAs. (m-o) consider static displacement scattering from lattice distortion, which causes weaker versions of the same intensities.
}
\end{figure*}

Electron DPs from this class of alloys consistently exhibit intensities that have been widely, though not universally, attributed to SRO.
Specific observations, which depend on the crystallographic zone axis (ZA) of incident electrons, are explicitly chronicled in Ref. \cite{walsh23a}.
As detailed in Methods, {Fig. \ref{fig:crconi}} reproduces the three main features of interest in a representative CrCoNi alloy: (a) streaking in the [110] ZA, (b) diffuse $\frac{1}{3}\{422\}$ superlattice intensities in the $[\bar{1}1\bar{1}]$ ZA, and (c) diffuse $\frac{1}{2}\{311\}$ superlattice intensities in the $[\bar{1}12]$ ZA.
As a reference, the diffractions expected from perfect lattices of random CrCoNi are provided in (d-f), as calculated using multislice simulations described in Methods.
Interestingly, (e) and (f) contain a few very faint `forbidden' diffractions, which are attributed to dynamical scattering that is discussed later.

The diffuse $\frac{1}{2}\{311\}$ intensities in the $[\bar{1}12]$ DP shown in (c) could, in principle, originate from previously proposed CuPt-type ($\mathrm{L}1_{1}$) ordering, but this form of SRO would also cause missing $\frac{1}{2}\{111\}$ diffractions \cite{walsh23a}, as well as equivalent intensities in other ZAs.
CuPt-type ordering is additionally inconsistent with the predictions of standard theoretical methods, as explicitly shown in {Supplementary Fig. 1}.
Further casting doubt on the observation of SRO, Ref. \cite{miller16} noted the existence of $\frac{1}{3}\{422\}$ intensities in pure Ni, while Ref. \cite{li23} more recently reported $\frac{1}{3}\{422\}$ and $\frac{1}{2}\{311\}$ diffractions in pure Cu.
Nonetheless, despite some speculation \cite{xu15,miller16,kawamura21,li23,walsh23a}, an alternative explanation for these results has heretofore not been established.
In this article, a combination of TEM techniques and diffraction simulations are used to demonstrate that static and thermal displacements can account for most previously reported observations, with certain results attributable to stacking faults, surface terminations, and dynamical scattering.

\section{Results}

\subsection{Extra diffractions from thermal displacements}
Thermal scattering \cite{vantendeloo98,wang03} was in fact one of the earliest interpretations of diffuse intensities in concentrated fcc alloys \cite{xu15}, though the explanation does not seem concretely established.
As detailed in Methods, diffraction simulations can approximate thermal excitations by displacing atoms from their ideal positions according to experimentally informed normal distributions.
Figures \ref{fig:crconi}(f-i) show that random deviations from perfect lattice sites cause relatively uniform diffuse scattering, but cannot account for specific extra diffractions.
More realistically, (j-l) present DPs calculated for snapshots of molecular dynamics (MD) simulations driven by a `machine learning' interatomic potential \cite{du22}, which are further described in Methods.
Some faint streaking appears in the [110] DP in (j), though this seems somewhat weaker than in, for example, Ref. \cite{zhang20}.
More clearly, calculations for the (k) $[\bar{1}1\bar{1}]$ and (l) $[\bar{1}12]$ ZAs predict DPs that closely resemble the experiments shown in (b) and (c).

Considering broad similarities in the phonon spectra of fcc metals \cite{antonov90}, the same phenomenon is expected to explain observations in other systems.
Indeed, {Supplementary Fig. 6} predicts comparable thermal scattering in pure Cu, excellently reproducing the DPs reported by Ref. \cite{li23}. 
Experimentally, thermal scattering can be isolated by considering, in addition to DPs, the Fourier transforms (FTs) of atomic resolution STEM images.
As directly imaged atomic positions represent time-averages, specific intensities from thermal scattering should disappear \cite{vantendeloo98}.
Accordingly, extra diffractions in, for example, the FTs of VCoNi images \cite{chen21} must have originated through other means.
Diffuse $\frac{1}{3}\{422\}$ intensities in the 77 K DP of a Ni-Cr-based alloy \cite{miller16} also likely have a distinct source, though Supplementary Fig. 6 suggests that thermal scattering may not be entirely negligible at this temperature.

\subsection{Extra diffractions from static displacements}

Possibly explaining these observations, Figs. \ref{fig:crconi}(n,o) demonstrate that the introduction of energetically favorable lattice distortion through structural relaxation (see Methods) produces weaker versions of the diffuse intensities seen in (k) and (l).
This static displacement scattering has been studied historically \cite{cook69b,vantendeloo98,ice99} and was previously proposed as a potential source of extra diffractions in concentrated fcc alloys \cite{zhou21,kawamura21}.
Equivalent CrNi$_2$ configurations, which are representative of the alloy considered in Ref. \cite{miller16}, cause similar displacement scattering, but extra diffractions are largely absent from calculations for CoNi$_2$, which forms a more ideal lattice on account of the chemical similarity of the two elements \cite{oh21}.
V, which electronegatively resembles Cr, is expected to cause comparable or greater scattering in VCoNi.
Applying random displacements to relaxed structures mostly adds background intensity, confirming that correlated thermal displacements contribute a separate, stronger effect.
Together, static and thermal displacement scattering can explain every measurement of which we are aware, except the FT of a HAADF STEM image of [111]-oriented Cu \cite{li23}, where the innermost $\frac{1}{3}\{422\}$ diffractions appear as sharp peaks.
(Similar analysis was not provided for the [112] ZA.)

\subsection{Extra diffractions from dynamical amplification}

It is not impossible that the extra diffractions in the FT of Cu \cite{li23} originated from hidden planar defects, which would be accommodated by the low SF energy of Cu and are examined in more detail later.
However, the faint intensities in Fig. \ref{fig:crconi}(e) indicate that the same extra diffractions can, in principle, occur in perfect lattices.
The nature of these features is inferred from calculations for varying-thickness Ni foils presented in Fig. \ref{fig:ni}.
Experimental DPs from a \SI{\sim120}{\nm} sample in the (a) $[\bar{1}1\bar{1}]$ and (b) $[\bar{1}12]$ ZAs are provided for reference.
Thermal scattering, which is significantly weaker in Ni than Cu \cite{peng96} or CrCoNi \cite{zhang17a}, is not readily visible in (b), though likely contributed to the extra diffractions in (a).
Still, these results provide a benchmark for theoretical calculations, which considered only random thermal excitations in order to isolate scattering from other sources.

Fig. \ref{fig:ni}(c) predicts that non-negligible $\frac{1}{3}\{422\}$ diffractions will always appear in the $[\bar{1}1\bar{1}]$ ZA of \SIrange{\sim15}{100}{\nm} foils.
The absence of significant intensities below \SI{\sim15}{\nm} suggests that they arise through multiple scattering events that increase in frequency with sample thickness, though this phenomenon is generally understood to modify existing intensities, not generate new diffractions \cite{vantendeloo98}.
Absent other sources, these features could ultimately originate from otherwise negligible higher order Laue zone (HOLZ) diffraction, that is, scattering between, rather than within, reciprocal lattice planes normal to the ZA, as was suggested by Ref. \cite{miller16}.
Specifically, in the $[\bar{1}1\bar{1}]$ ZA, kinematically miniscule $(11\bar{1})$ diffractions from the immediately above plane of the reciprocal lattice would be projected onto $\frac{1}{3}(42\bar{2})$, while $(1\bar{1}\bar{1})$ projections from the plane below would appear as $\frac{1}{3}(2\bar{2}\bar{4})$, etc.
Reduced intensities above \SI{\sim100}{\nm} would be consistent with the decay of underlying HOLZ diffractions with increasing sample thickness.
While this dynamical scattering explains the faint peaks in Figs. \ref{fig:crconi}(e,f,h,i), it is not expected to majorly contribute to the DP of the \SI{\sim120}{\nm} foil displayed in Fig. \ref{fig:ni}(a).

Figure \ref{fig:ni}(d) provides equivalent calculations for the $[\bar{1}{1}2]$ ZA, where (111) diffractions in the above plane of the reciprocal lattice could be projected to $\frac{1}{3}(421)$, while $(11\bar{1})$ in the plane below could appear at $\frac{1}{3}(24\bar{1})$, etc.
These locations are marked in (b), though no such scattering is visible experimentally, consistent with the comparably minimal intensities predicted in (d).

\subsection{Extra diffractions from surface terminations}

\begin{figure*}
\includegraphics[width=172mm]{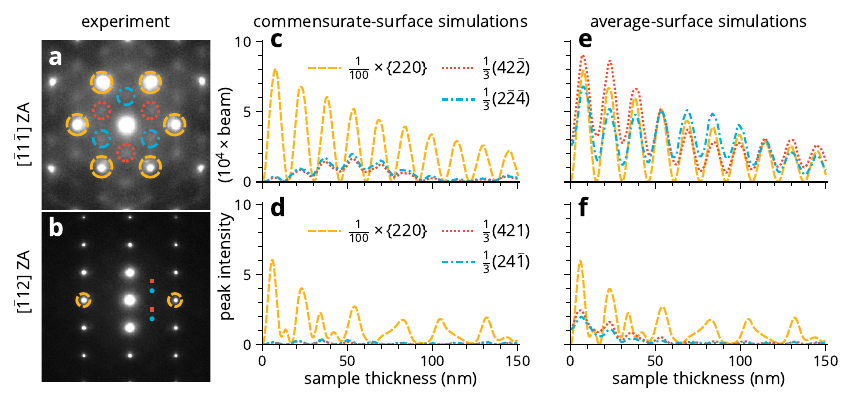}
\caption{\label{fig:ni}
\textbf{Extra diffractions from dynamical scattering and incommensurate surface terminations in Ni.}
Experimental DPs from a $\SI{\sim120}{\nm}$ Ni foil in the (a) $[\bar{1}1\bar{1}]$ and (b) $[\bar{1}12]$ ZAs.
Annotations indicate $\{220\}$ lattice diffractions (yellow-dashed) and locations where two sets of extra diffractions can theoretically occur (red dotted and blue dot-dashed).
See Fig. \ref{fig:crconi} for coordinates.
(c,d) Peak intensities at the highlighted locations in simulations of periodic Ni configurations with random thermal displacements.
Reference $\{220\}$ intensities are plotted at one hundredth scale. 
Clear extra diffractions are predicted in (c), if not (d).
(e,f) are equivalent to (c,d) for simulations considering all possible surface terminations.
Each datum represents an average of (e) three or (f) six possible stacking multiples, which are predicted to cause the same extra diffractions that occur in commensurate configurations.
}
\end{figure*}

As discussed in Methods, all structures simulated up to this point have been periodic.
However, a complete investigation must account for samples in which the number of atomic layers along the ZA is not a multiple of the bulk periodicity.
For example, a [111]-oriented foil can consist of $3m$ commensurate layers, where $m$ is an integer, or $3m\pm1$ incommensurate layers, which are known to produce $\frac{1}{3}\{422\}$ intensities, plus reciprocal lattice translations.
Similarly, there exist six possible planar terminations of [112] and two of [110].
As more recently noted by Refs. \cite{miller16,li23}, Ref. \cite{cherns74} demonstrated this phenomenon in Au films deposited in (111) layers, for which dark-field images reveal surface contours corresponding to atomic steps.

In conventional jet-polished TEM samples, dark-field apertures around diffuse intensities highlight nm-scale features that have been previously interpreted as ordered domains.
Following Ref. \cite{cherns74}, these could instead correspond to tiny surface islands of varying atomic thickness, in which case standard DPs sampling hundreds of \SI{}{\nm^2} would simultaneously represent many different commensurate and incommensurate stacking terminations.

Calculations for all types of surface termination are individually presented in Supplementary Fig. 7 and averaged in Fig. \ref{fig:ni} for the (e) $[\bar{1}1\bar{1}]$ and (f) $[\bar{1}12]$ ZAs, analogous to commensurate calculations in (c,d).
Interestingly, incommensurate surface terminations generate the same extra diffractions, though with much greater intensity.
Calculations for the $[\bar{1}1\bar{1}]$ ZA in Fig. \ref{fig:ni}(b) predict $\frac{1}{3}\{422\}$ diffractions of magnitude not incomparable to experimental values shown in (a).
While the simulated features are sharper than in ambient temperature experiments, which presumably include thermal scattering, the demonstrated athermal effect could contribute to low-temperature and real-space observations.
Intriguingly, calculations suggest that the diffuse intensities could split into two groups corresponding to $\frac{1}{3}(42\bar{2})$ and $\frac{1}{3}(2\bar{2}\bar{4})$, plus reciprocal lattice translations---this has not, to our knowledge, been observed experimentally.
The extra diffraction predicted for the $[\bar{1}12]$ ZA in (f) remain small, though calculations indicate that intensities could become significant in very thin samples. 
As shown in Supplementary Fig. 7, this mechanism does not explain observations in the [110] ZA.

\subsection{Extra diffractions from planar defects}
\begin{figure*}
\includegraphics[width=172mm]{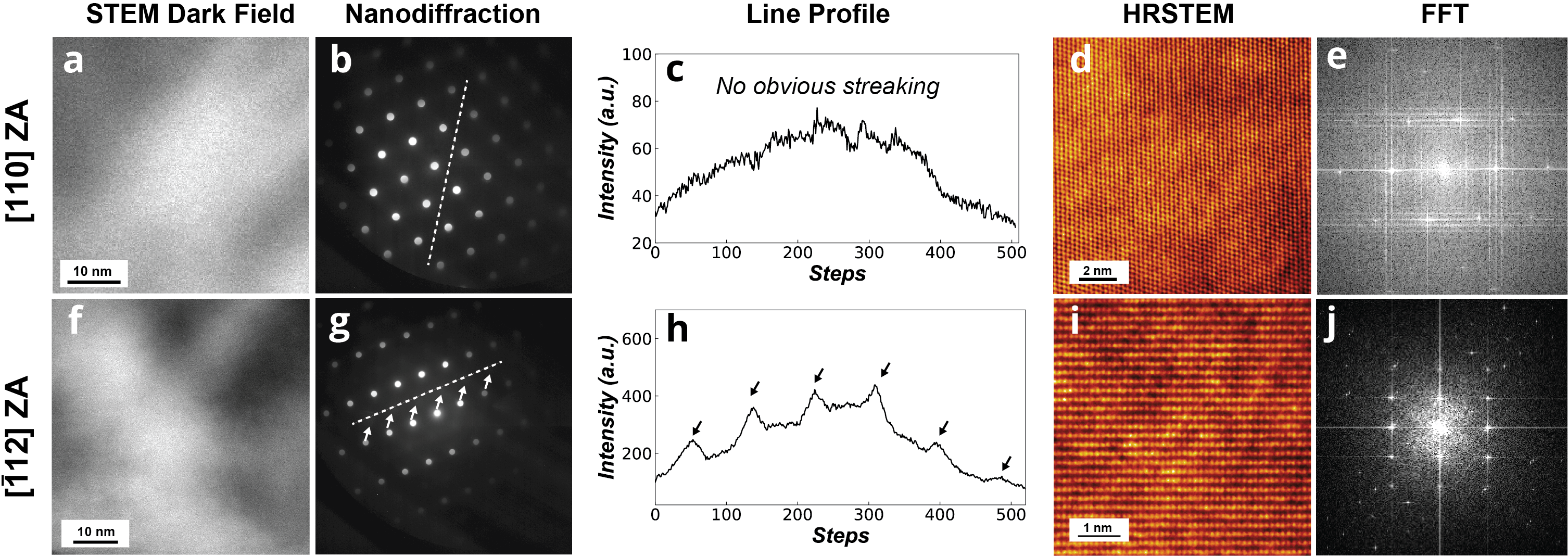}
\caption{\label{fig:slowcool}
    \textbf{Absence of planar defects in slow-cooled CrCoNi samples.} (a) STEM-DF image of a \SIrange{20}{30}{\nm} thick region (down-to-up) in a slow-cooled CrCoNi sample along the [110] ZA; (b) summed nanoDP from (a); (c) the intensity profile of the dashed line marked in (b), showing no obvious streaking; (d) atomic-resolution STEM-LAADF image showing no SFTs; (e) the Fourier transform of (d), showing no streaking along $\langle111\rangle$ g-vectors; and (f-j) equivalent to (a-e) for the same region in the $[\bar{1}12]$ ZA. Despite the absence of defects, the line profile given in (h) shows clear diffuse $\frac{1}{2}\{311\}$ superlattice peaks, as indicated by the black arrows.
}
\end{figure*}

While previously considered phenomena can fully account for results in the [111] and [112] ZAs, it is worth noting that all the features of interest can also be caused by stacking faults (SFs), which are usually readily visible in TEM, particularly when using bright-field (BF) or dark-field (DF) imaging modes with diffraction contrast.
Conventional microscale faulting can explain recently reported ``mechanically-driven SRO" \cite{seol22}, where extra diffractions likely originated from SFs formed between Shockley partials dissociated across slip bands induced by mechanical deformation.
The creation of Frank loops, which enclose stacking faults, by ion irradiation \cite{hull11} can similarly account for recently proposed ``irradiation-assisted SRO" \cite{su22}.
Besides these apparent SFs caused by mechanical deformation or ion irradiation, conventional sample preparation methods, namely Ar ion milling and Ga focused ion beam (FIB) milling, can inadvertently introduce Frank loops with diameters as small as \SIrange{2}{5}{\nm} on foil surfaces, as shown in Supplementary Figs. 3(g,h).
This type of defect could affect results in ion-milled samples \cite{picak23,tang23}, but most TEM specimens, including those examined in this study, are instead prepared only by electropolishing, which does not introduce structural imperfections.

We have previously speculated \cite{walsh23a} that less obvious nanoscale stacking fault tetrahedra (SFTs) or Frank loops could nonetheless form during bulk sample processing.
Consistent with this hypothesis, {Supplementary Figs. 4 and 5} reveal a remarkable abundance of previously unreported SFTs in water-quenched CrCoNi.
However, streaking in the [110] ZA is not widespread, while $\frac{1}{2}\{311\}$ diffractions are equally present in slow-cooled samples, which do not contain quench-induced faulting.
The absence of planar defects in slow-cooled samples is demonstrated in Fig. \ref{fig:slowcool}, which presents (a,f) DF and (d,i) HRSTEM images for the [110] and $[\bar{1}12]$ ZAs of the same region.
While nanoscale faulting can locally cause streaking, Figs. \ref{fig:slowcool}(b,c) shows that the feature is often absent from larger scale DPs.
In contrast, (g,h) indicate that a $\frac{1}{2}\{311\}$ superlattice persists independent of processing conditions.
Defects smaller than \SI{2}{\nm} could theoretically escape detection by HR(S)TEM, but such features would not be expected to contribute to typical DPs.
It remains possible that specific observations of streaking in the $[110]$ ZA reflect larger, more heterogeneously distributed planar defects.

\section{Discussion}

Building on Ref. \cite{miller16}, Ref. \cite{coury23} very recently proposed that the HOLZ diffraction examined in Fig. \ref{fig:ni}(c,d) is the common source of widely observed diffuse intensities.
However, it is not clear if the phenomenon can be generally distinguished from theoretically stronger surface contributions examined in (e,f).
Of course, these effects are greatly complicated by the true conditions of foil surfaces, which, for the considered alloys, inevitably host nm-scale oxides.
Analysis is further obfuscated by the reality that TEM ZAs rarely correspond to actual surface normals with associated steps, but are instead obtained by tilting grains with imprecisely known alignments.
Imperfect experimental correspondence questions the realism of the present calculations, but we do not believe that surface effects can be altogether dismissed on the basis of existing evidence.

In any case, dynamically amplified HOLZ diffraction seems less significant outside of the [111] ZA.
Indeed, contrary to prior assertions \cite{miller16}, this mechanism does not produce experimentally observed $\frac{1}{2}\{311\}$ intensities in the [112] ZA, rather tiny peaks at positions such as $\frac{1}{3}(421)$, which are identified in Fig. \ref{fig:ni}(b).
Ref. \cite{coury23} provides nominally correct projections in the $[112]$ ZA, but these are not plotted in exactly the right positions---sharp HOLZ diffractions lie outside, not within, many observed features.
It is thus unclear how adjacent $\frac{1}{3}\{421\}$ diffractions could generate $\frac{1}{2}\{311\}$ intensities, though they could be subsumed into diffuse scattering from other sources, which occurs to an extent in Figs. \ref{fig:crconi}(l,o).
Still, the proximity of displacive and HOLZ intensities is not necessarily coincidental, as local lattice distortions and thermal excitations could, in principle, generate intensities via out-of-plane scattering.
Detailed analysis of possible connections is beyond the scope of this study, which aims to practically explain observations in the literature.

As a final digression, it is worth emphasizing the extent to which both lattice and diffuse intensities vary with sample thickness, as seen in Fig. \ref{fig:ni}.
This effect should be less extreme in real experiments, in which electrons are not ideally monochromatic and samples vary in height, but thickness differences can still compromise comparison between any two specimens.
For example, the absence of clear diffuse intensities in one observation of an unaged alloy implies relatively little about the origin of intensities in an aged alloy, particularly when lattice diffractions in the unaged DP are also considerably weaker \cite{liu23}.
This is to say nothing of local variations in surface topography and oxides.
Differences in intensities reported in our earlier work \cite{zhang22} could similarly be explained by variations in sample details.

\section{Conclusion}
Diffuse intensities previously attributed to SRO are caused by a number of phenomena generally associated with the breaking of lattice symmetry.
The greatest source of this scattering appears to be correlated, though not random, thermal displacements, which likely contributed to most reciprocal-space observations in the [111] and [112] ZAs.
Intensities in FTs and low-temperature DPs of concentrated alloys could instead represent static displacement scattering, which qualitatively produces the same features.
Extra diffractions in the [111] ZA can also be caused by incommensurate surface terminations, though further work is needed to understand these effects in realistic samples.
Dynamical artifacts also appear at the same locations, but to a lesser extent that seems insufficient to explain most observations.

In some studies, extra diffractions may primarily reflect planar defects, though in most cases these were otherwise apparent.
While a surprising abundance of nearly invisible SFTs was found in quenched CrCoNi, these appear not to affect larger scale DPs and were not observed in slow-cooled samples.
There is relatively little evidence that electron diffuse intensities directly relate to SRO, which may still widely exist in concentrated 3\textit{d} alloys.
It could be interesting for future studies to clarify if variations in SRO can cause detectable changes in static and thermal displacement scattering.

\section{Methods}

\subsection{Specimen processing}
Equiatomic CrCoNi alloys were produced by arc-melting high-purity ($>99.9$\%) elements under an Ar atmosphere followed by drop casting into $\SI{127}{\mm}\times\SI{19.1}{\mm}\times \SI{25.4}{\mm}$ Cu molds.
As-cast samples were homogenized in vacuum at \SI{1200}{\celsius} for \SI{24}{h} and cold-rolled to a final thickness of \SI{6.1}{\mm} (76\% reduction).
Samples were then recrystallized at \SI{1000}{\celsius} for \SI{0.5}{h}, followed by either ice water quenching or slow furnace cooling.

\subsection{Transmission electron microscopy}
Energy-filtered selected area electron DPs were obtained on a Zeiss monochromated LIBRA 200MC microscope at an accelerating voltage of 200 kV in the [110], $[\bar{1}1\bar{1}]$, and $[\bar{1}12]$ zone axes. Inelastically scattered electrons were filtered by an in-column $\Omega$ energy filter using an energy slit of \SI{5}{eV}. While thermal scattering is technically inelastic, electron energies are only minimally affected by interactions with phonons \cite{vantendeloo98}, which typically have energies below \SI{1}{eV}, so thermal scattering should not be removed by standard filters.

STEM imaging of quenched and slow-cooled CrCoNi was performed on the Transmission Electron Aberration-corrected Microscope (TEAM) I at the National Center for Electron Microscopy at Lawrence Berkeley National Laboratory. A wide range of microscope parameters in terms of  C2 aperture sizes, convergence angles ($\alpha$), current density, and camera lengths (CL) was explored to acquire optimal images for each imaging technique. Diffraction-contrast bright- and dark-field STEM imaging was carried out under $\vec{g} = 200$ two-beam conditions near the [110] ZA using a \SI{5}{\um} C2, an $\alpha$ of \SI{1.1}{mrad}, and CLs from \SI{2.2}{m} to \SI{4.3}{m}. Scanning nanobeam diffraction, or 4-dimensional STEM (4DSTEM) was performed using a 5 or \SI{10}{\um} C2 and resulting $\alpha$ of 1.1 or \SI{2.3}{mrad} and probe sizes of \SI{1}{nm} and \SI{0.7}{nm}, and CLs from \SI{0.63}{m} to \SI{0.8}{m} in [110] and $[\bar{1}12]$ ZA on a Gatan K3 camera. Since a tradeoff exists between real space and reciprocal space resolution, the convergence angle was selected between these two values based on an \textit{ad hoc} basis. Inelastically scattered electrons were filtered by an post-column Gatan Imaging Filter (GIF) Continuum K3 spectrometer using an energy slit of \SI{20}{eV}. Raw 4DSTEM data were processed by the py4DSTEM package \cite{savitzky21}.
High-resolution STEM was conducted in the same locations as in 4DSTEM scans using a \SI{70}{\um} C2 and an $\alpha$ of 16 mrad. A large CL of \SI{0.8}{m} was used to ensure the images were taken under low-angle annular dark-field (LAADF) to capture the diffraction contrast originating from crystal defects. Conventional high-angle annular dark-field (HAADF) was not used because CrCoNi provides very limited Z-contrast.
The probe was corrected using a CEOS DCOR spherical-aberration corrector to an outer tableau tilt of \SI{35}{mrad} and C1, A1, A2, B2, C3, A3, S3, C5 values of \SI{-2.5}{nm}, \SI{2.4}{nm}, \SI{15.5}{nm}, \SI{20.0}{nm}, \SI{103.8}{nm}, \SI{96.8}{nm}, \SI{104.4}{nm}, and \SI{-771}{\um}, respectively.

TEM samples were prepared by the following steps: mechanical polishing to a final thickness of \SI{50}{\um} using 320/600/800/1200 grit papers; punching out \SI{3}{mm} discs from the foils and dimpling on one side to a center thickness of \SI{\sim20}{\um}, and eventually achieving perforation by twin-jet polishing on a Fischione Model 110 electropolisher in an electrolyte of 70\% methanol, 20\% glycerol, and 10\% perchloric acid at \SI{-30}{\celsius} under a stable current of \SI{\sim25}{mA}. Significant care has been taken to ensure that the thinnest part of the jet-polished samples was \SIrange{20}{30}{\nm}, which was confirmed by the electron energy loss spectroscopy (EELS) log-ratio technique.

In order to examine possible artifacts caused by argon ion milling, selected jet-polished samples were further ion-milled by a Gatan PIPS II precision ion-polishing system under a voltage of \SI{3}{kV} for \SI{1}{h}, followed by \SI{1}{kV} for \SI{15}{min}, and \SI{0.5}{kV} for \SI{30}{min}. The guns were aligned to \SI{\pm3}{\degree}. Alternatively, focused ion beam (FIB)-liftout samples were prepared on an FEI Scios 2 DualBeam FIB/SEM under an operating voltage of \SI{30}{kV} followed by polishing at \SI{5}{kV} and \SI{2}{kV}. The FIB lamellae were then polished on a Fischione Model 1040 Nanomill under voltages of 900 and \SI{500}{V} for \SI{20}{min} to further alleviate FIB damage.  The ion-milled and FIB’ed samples also have a final thickness of \SIrange{\sim20}{30}{\nm}. These samples were characterized by an FEI F20 UT Tecnai STEM under an accelerating voltage of \SI{200}{\kV}.

\subsection{Diffraction simulations}
Multislice diffraction simulations were performed using abTEM \cite{madsen21} with atomic scattering potentials parameterized by Ref. \cite{lobato14}.
For the calculations displayed in Fig. \ref{fig:crconi}, potentials were projected onto \SI{0.25}{\angstrom} slices using finite integrals and a sampling resolution of \SI{0.02}{\per\angstrom}.
All simulation cells were $\SI{10}{\nm} \times \SI{10}{nm}$ in directions normal to the ZA, with varying thicknesses.
CrCoNi configurations were \SI{\sim27.5}{\nm} thick, with the goal of representing \SIrange{\sim20}{30}{\nm} samples experimentally considered in this study and the \SI{\sim30}{\nm} samples considered by Ref. \cite{chen21}.
Structures for the $[110]$, $[\bar{1}1\bar{1}]$, and $[\bar{1}12]$ ZAs respectively contained 218, 135, and 378 atomic layers along the ZA.
While a comprehensive thickness study with realistic displacements is hardly feasible, this height seems to represent average scattering conditions in the vicinity, with stronger dynamical effects calculated for \SI{\sim25}{\nm} configurations and weaker effects calculated for \SI{\sim30}{\nm} configurations, consistent with the findings displayed in Fig. \ref{fig:ni}.
These simulations are intended to qualitatively demonstrate the discussed scattering mechanisms, not as explicitly quantitative predictions.

Due to the large number of simulations involved, calculations for Fig. \ref{fig:ni} used a slightly more approximate \SI{0.5}{\angstrom} slice thickness, \SI{0.05}{\per\angstrom} sampling, and infinite integral projections.
The difference in parameters was found to only minimally affect calculations for finite-temperature configurations, though in some cases even larger pixel sizes were noted to exaggerate extra diffractions.
All calculated DPs are linearly plotted over the intensity range 0 to $1.2\times10^{-5}$, where the original beam intensity is 1, using the matplotlib \cite{hunter07} colorscheme ``binary\_r."
Experimental DPs are plotted similarly, albeit without definitive intensity normalization.
As the assumption of ideally monochromatic plane-waves in multislice simulations produces unrealistically sharp DPs, a Gaussian filter with a width of 2 pixels was applied to all plotted images, which slightly diffuses, though in sum preserves, calculated intensities.
These consistent settings were chosen to approximately reproduce the lattice diffractions in comparable experiments, while also clearly showing various diffuse intensities.
It should be noted that this procedure somewhat reduces the visibility of dynamical artifacts, which are typically single pixels, but this effect is expected to be realistic and may explain why intensities such as $\frac{1}{3}\{421\}$ are not seen in experiments.
All quantitatively reported intensities represent local maxima before filtering.

Random phonons were modeled using 16 configurations with normally distributed displacements with experimentally informed standard deviations of \SI{0.06714}{\angstrom} for Ni \cite{peng96} and \SI{0.0922}{\angstrom} for CrCoNi \cite{zhang17a}. The equivalent value for Cu, which was not used, is \SI{0.08935}{\angstrom} \cite{peng96}.

\subsection{Molecular statics and dynamics}
Static structural relaxations and MD simulations were performed using LAMMPS \cite{thompson22}. 
In order to model \SI{0}{\kelvin} lattice distortion, atomic positions were energetically optimized following a conjugate-gradient algorithm until the collective norm of all forces was below \SI{e-12}{eV/\angstrom}.
Realistic phonons were obtained by averaging 16 MD snapshots, which were taken every 2000 steps after 10000 steps of equilibration, where each step is \SI{1}{fs}.

For computational simplicity, simulations employed fixed lattice parameters according to experimental values at ambient conditions, that is \SI{3.524}{\angstrom} for Ni, \SI{3.615}{\angstrom} for Cu \cite{haynes16}, and \SI{3.56}{\angstrom} for CrCoNi \cite{jin17}.
Constant-pressure simulations were found to produce equivalent DPs.
CrCoNi was modeled using the neural network potential \cite{singraber19} described in Ref. \cite{du22}. It should be noted that while this model seems to reasonably represent underlying density-functional theory calculations, a widely used embedded-atom method potential \cite{li19}, which does not realistically describe atomic interactions \cite{ghosh22}, predicts structures without significant static displacement scattering. Ni was modeled only as an ideal lattice with random thermal displacements.

Both periodic and free surfaces in the direction of the ZA were considered.
While the latter may seem more realistic, the CrCoNi potential displayed some instabilities in the presence of free surfaces.
Additionally, MD simulations of surfaces exhibited unphysically large swaying motions that were likely related to the lack of constraint and imposition of periodic boundary conditions perpendicular to the ZA.
These could be eliminated through careful optimization of the thermostat, but DPs were otherwise identical to fully periodic calculations, which were used instead for convenience.

\begin{acknowledgments}
  This work was supported by the US Department of Energy, Office of Basic Energy Sciences, Materials Sciences and Engineering Division under contract No. DE-AC02-05CH11231 as part of the Damage-Tolerance in Structural Materials (KC13) program. Simulations were performed using the Lawrencium computational cluster provided by the IT Division of Lawrence Berkeley National Laboratory (supported by the same office and contract number), as well as award No. BES-ERCAP0021088 of the National Energy Research Scientific Computing Center, a US Department of Energy Office of Science User Facility operated under the same contract number.
  C. Ophus is thanked for advice on diffraction simulations.
\end{acknowledgments}

\end{document}